\documentclass[prb,aps,twocolumn]{revtex4}
\usepackage{amsmath}
\usepackage{amssymb}
\usepackage {graphicx}

\newcommand{\beq}{\begin{equation}}
\newcommand{\eeq}{\end{equation}}
\newcommand{\bea}{\begin{eqnarray}}
\newcommand{\eea}{\end{eqnarray}}
\newcommand{\e}{\varepsilon}

\newcommand{\bk}{{\vec k}}
\newcommand{\bp}{{\vec p}}
\newcommand{\bq}{{\vec q}}

\newcommand{\nn}{\nonumber}
\newcommand{\bse}{\begin{subequations}}
\newcommand{\ese}{\end{subequations}}
\newcommand{\bwt}{\begin{widetext}}
\newcommand{\ewt}{\end{widetext}}

\newcommand{\bkp}{{\vec k}'}

\newcommand{\I}{\mathrm{Im}}
\newcommand{\R}{\mathrm{Re}}
\newcommand{\bsu}{\begin{subequations}}
\newcommand{\esu}{\end{subequations}}

\begin{document}

\title{Intrinsic Damping of Collective Spin Modes\\
in a Two-Dimensional Fermi Liquid with Spin-Orbit Coupling}
\author{Saurabh Maiti$^{1,2}$ and Dmitrii L. Maslov$^1$}
\affiliation {~$^1$Department of Physics, University of Florida, Gainesville, FL 32611}
\affiliation {~$^2$National High Magnetic Field Laboratory, Tallahassee, FL 32310}
\date{\today}

\begin{abstract}
A Fermi liquid with spin-orbit coupling (SOC) is expected to support a new kind of collective modes: oscillations of magnetization in the absence of the magnetic field. We show that these modes are damped by the electron-electron
interaction even in the limit of an infinitely long wavelength ($q=0$). The linewidth of the collective mode
is on the order of $\bar\Delta^2/E_F$, where $\bar\Delta$ is a characteristic spin-orbit energy splitting and $E_F$ is the Fermi energy. Such damping is in a stark contrast to known damping mechanisms of both charge and spin collective modes in the absence of SOC, all of which disappear at $q=0$, and arises
because none of the components of total spin is conserved in the presence of SOC.
\end{abstract}

\maketitle

Electron systems with spin-orbit coupling (SOC) exhibit rich physics, some of which
may have technological applications; \cite{spintronics1,spintronics2} equally rich is the physics of cold-atom systems with synthetic SOC.\cite{ca1,ca2} Combining many-body interactions with broken SU(2) symmetry, one obtains a special--``chiral''--kind of Fermi liquid (FL)\cite{1,ali_rashba_maslov,2}  that supports a new type of collective modes, ``chiral-spin waves''--oscillations of the spin density in zero magnetic field.\cite{shekhter,ali_maslov,Zhang,sm}

In the absence of SOC, electron-electron interaction ({\em eei})\cite{footnote_eei} does not affect certain properties of an electron system given that some symmetries are preserved. For example, the conductivity and cyclotron-resonance frequency of a Galilean-invariant system are not affected by {\em eei}; same is true for the de Haas-van Alphen (dHvA) frequency in an isotropic system\cite{comment_dhva} and  for the Larmor frequency in the presence of an SU(2)-symmetric interaction. Being a relativistic effect, SOC breaks both Galilean (but not necessarily rotational) invariance and SU(2) symmetry and thus lifts the protection ensured by these symmetries. As a result, several physical quantities become dependent on {\em eei}. The list of such quantities includes the optical conductivity,\cite{Mishchenko_Farid1} Drude weight,\cite{Agarwal}, and frequencies of collective spin modes, which play the role of Larmor frequencies in zero magnetic field.\cite{shekhter,ali_maslov,Zhang,sm}

In this Letter, we discuss another fundamentally new effect induced solely by SOC: intrinsic damping of collective spin modes in the uniform ($q=0$) limit. Interaction-induced damping of collective modes is not, by itself, a new effect. For example, plasmons in 3D,\cite{kivelson:1969} 2D,\cite{reizer:2004} and 1D (Ref.~\onlinecite{Glazman}) electron systems, the Silin-Leggett (SL) mode\cite{silin:1958,leggett:1970} in a partially spin-polarized FL,\cite{ma:1968,mineev} and magnons in a ferromagnetic FL\cite{kondratenko:1976,mineev} are all
damped by interaction processes involving excitations of multiple particle-hole pairs. (This mechanism is different from Landau damping which involves only a single particle-hole pair: damping by multiple  pairs occurs even outside the single-particle continuum.)
However, Galilean invariance, in the case of charge modes,\cite{pines} and conservation of the total spin component along the field ($S_3$), in the case of spin modes,\cite{kondratenko:1965} ensure that this kind of damping vanishes at $q=0$. We show here that this is not the case for electron systems with SOC.

\begin{figure}[htp]
$\begin{array}{cc}
\includegraphics[width=0.4\columnwidth]{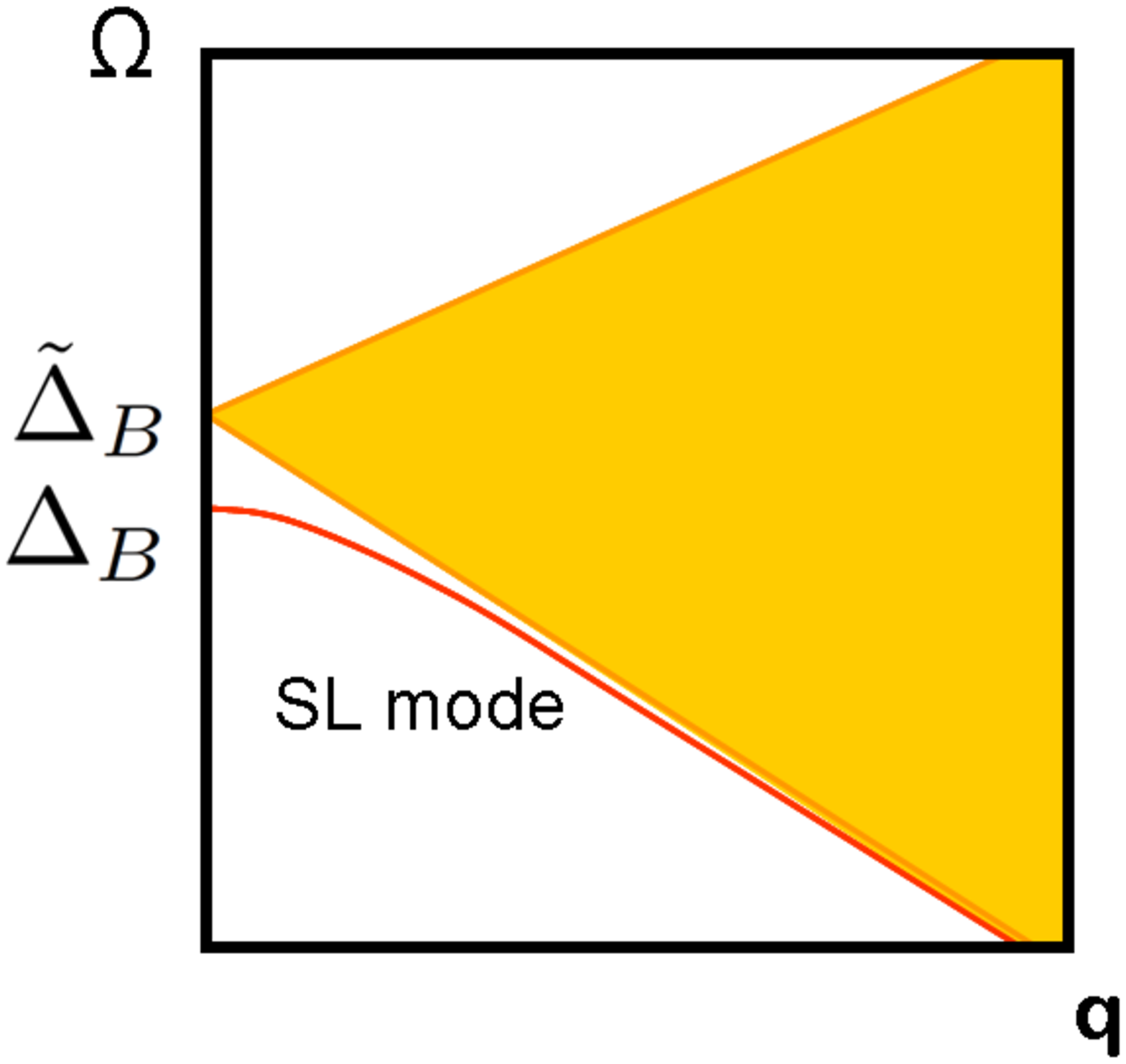}&
\includegraphics[width=0.42\columnwidth]{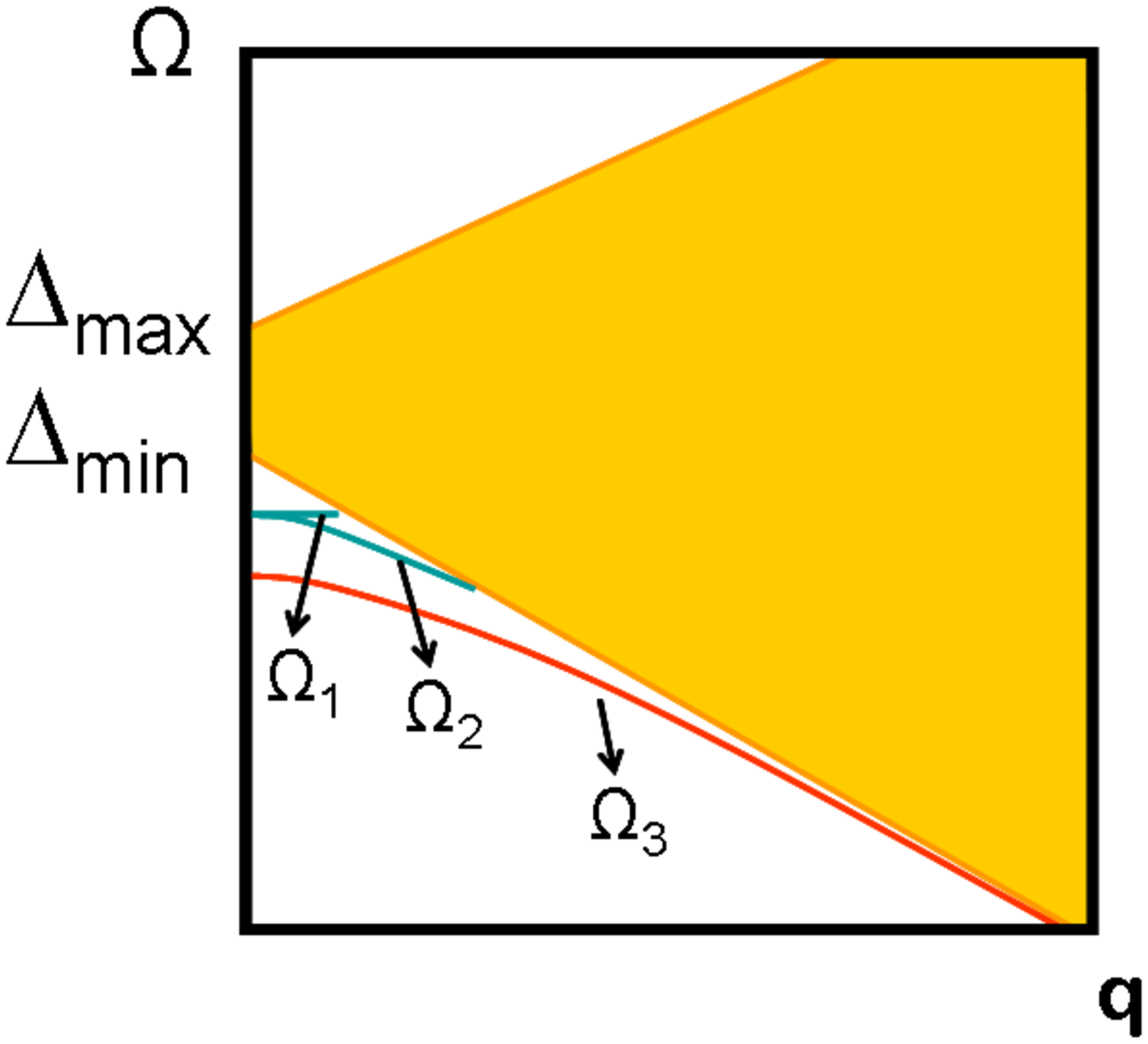}
\end{array}$\caption{
\label{fig:1} Left: Schematics of the Silin-Leggett mode in a partially spin-polarized
FL. Right: The chiral-spin modes in a FL with Rashba
spin-orbit coupling. The shaded regions denote the particle-hole continua, $\Delta_B$ is the Larmor frequency, $\tilde{\Delta}_B$ is the quasiparticle Zeeman energy, and $\Delta_{\min/\max}$ is the lower/upper boundary of the continuum at $q=0$.}
\end{figure}

\begin{figure}[htp]
$\begin{array}{c}
\includegraphics[width=0.9\columnwidth]{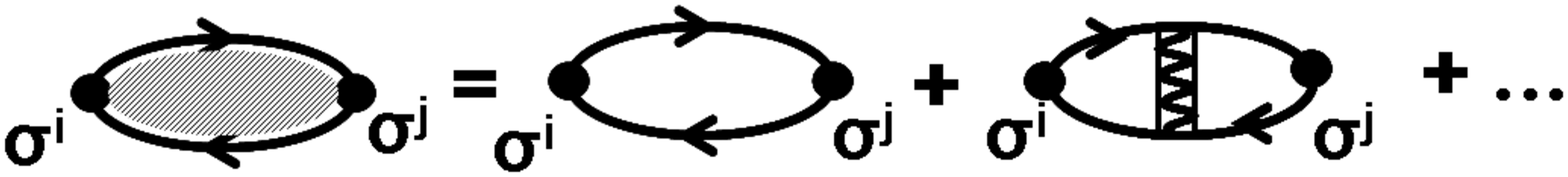}\\
\includegraphics[width=0.8\columnwidth]{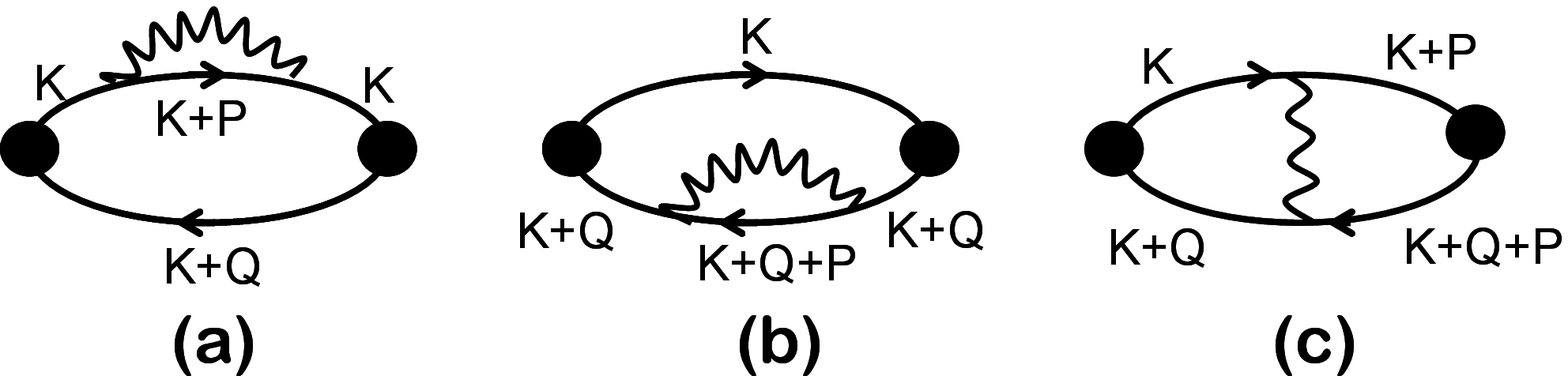}\\
\includegraphics[width=0.6\columnwidth]{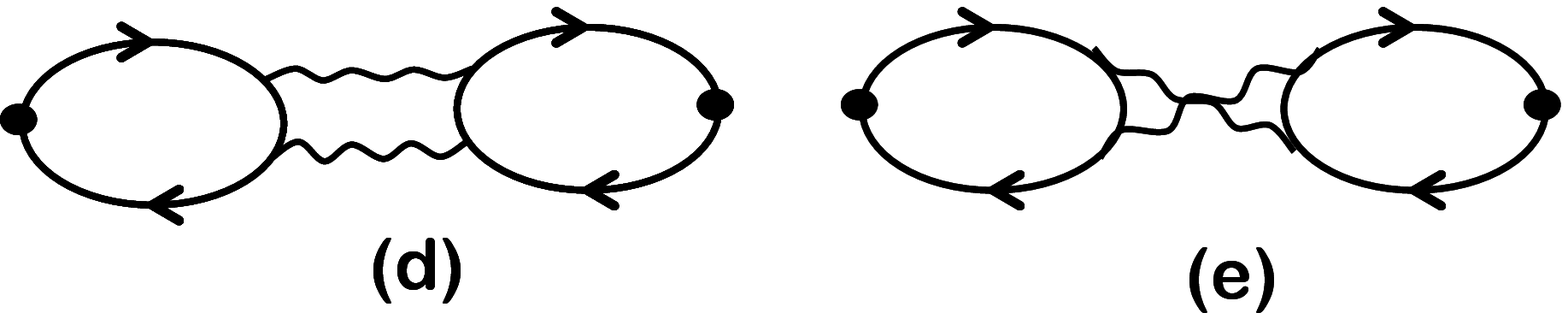}
\end{array}$\caption{ Top: The ladder (RPA) series for the spin susceptibility. The boxed wavy line is the static effective interaction $U_{\mathrm{x}}$. Bottom: Diagrams contributing to damping of the collective modes. The wavy line in diagrams {\em a-e} denotes a dynamic interaction, $V_{\mathrm{eff}}(P)$. \label{fig:3}}
\end{figure}

Two-dimensional (2D) electron systems with momentum-dependent SOC, e.g., of Rashba or Dresselhaus types, bear certain similarity to a partially spin-polarized Fermi gas. The latter has a transverse SL mode in the spin sector (see Fig.~\ref{fig:1}, left),\cite{silin:1958,leggett:1970} while the former has three (two transverse and one longitudinal) chiral-spin modes ($\Omega_{1}\dots \Omega_{3}$ in Fig.~\ref{fig:1}, right), \cite{shekhter,ali_maslov,Zhang,sm} which correspond to oscillations of the three components of magnetization. Indices $1-3$ label the Cartesian system with the $3$ axis along the normal to the plane of  a 2D electron gas. (Although SL-mode has been studied previously in 3D, the same mode should occur in 2D as well.) Conservation of $S_3$ ensures that the frequency of the SL mode at $q=0$ coincides with the Larmor frequency in the absence of {\em eei}. In the presence of SOC, none of the three spin components is conserved. As a result, one obtains three distinct modes with frequencies renormalized by {\em eei}. In addition, as we show here, these modes have finite linewidth which, in order of magnitude, is given by the inverse transport lifetime of a quasiparticle with energy equal to the spin-orbit splitting. That the SL mode at $q=0$ is not affected by {\em eei} follows already from the exact equations of motion for magnetization.\cite{ma:1968}
Diagrammatically, this occurs due to a cancellation between the self-energy and vertex graphs for the spin susceptibility.\cite{ma:1968} Such a cancellation, however, does not occur in the presence of SOC.

The single-particle Hamiltonian of a 2D system with SOC can be written as (we set $\hbar=1$) \bea\label{eq:a}
\mathcal{H}_k&=& \left(\frac{k^2}{2m}-\mu\right) \sigma_0
+\lambda \vec{\sigma}\cdot \vec{f}(\vec k),
\eea
where $\mu$ is the chemical potential, $\sigma_0$ is the $2\times 2$ unit matrix, $\vec{\sigma}$ is the three-dimensional vector of Pauli matrices, $\lambda$ is the SOC constant,  and $\vec{f}(\vec k)=-\vec f(-\vec k)$ is a 2D vector that depends on the details of SOC; e.g., $\vec{f}=(k_2,-k_1,0)$ for linear Rashba SOC.
The single-particle Green's function is given by
\beq\label{eq:Greens Function} G(K)=\sum_s{\Omega}_s(\bk)g_s(K), ~~{\Omega}_s(\bk)=\frac{1}{2}\left[\sigma_0+s
\hat\eta_{\vec k} \right], \eeq
where $g_s(K)=(ik_0-\frac{k^2}{2m}-s\Delta_{\vec k}/2+\mu)^{-1}$, $K\equiv(ik_0,\bk)$, $s=\pm1$ labels either spin projection or chirality, $\hat\eta_{\vec k}\equiv \vec{\sigma}\cdot\vec f/|\vec f|
$, and $\Delta_{\vec k}= 2\lambda |\vec f|$ is the spin-orbit splitting which, in general, depends not only on the magnitude but also on the direction of $\vec k$. We will be primarily interested in the case of weak SOC, when $\Delta_{\bk}\ll E_F$ for any $\bk$. In what follows, we will be comparing the $\Omega_3$ chiral-spin mode  to the (2D) SL mode, as both modes are transverse to the SOC-induced/Zeeman magnetic field. The latter can be described by the same Hamiltonian with $\vec{f}=(0,0,\Delta_B/2\lambda)$, where $\Delta_B\equiv g\mu_B B$, $g$ is the effective $g-$factor, $\mu_B$ is the Bohr magneton,  and $\vec B$ is the magnetic field chosen to be along the $3$-direction. The orbital effect of the field is not considered here.

Within the Random Phase Approximation (RPA), the spin susceptibility tensor is given by the ladder series in Fig.~\ref{fig:3}, where the boxed wavy line is a short-range interaction, $U_{\mathrm{x}}$, which mimics the exchange interaction in the spin channel. As shown in Ref.~\onlinecite{sm},
the frequencies of the collective modes correspond to the roots of the equation Det$\left(\sigma_0\otimes\sigma_0 +\frac{U_{\mathrm{x}}}{2}{\Pi}^0\right)$, where the elements of the $4\times 4$ spin-charge polarization matrix $\Pi^0$ are given by
\bea\label{eq:susceptibilities2}
\Pi^0_{ij}(Q)&=&\int_K \text{Tr}\left[\sigma_i G(K)\sigma_j G(K+Q)\right],\eea where
$\int_K\equiv T\sum_{k_0}\int\frac{d^2 k}{(2\pi)^2}$,
$i,j\in{0,1,2,3}$, and $0$ corresponds to the charge component. At $q=0$, all mixed spin-charge susceptibilities,
$\Pi^0_{0j}$ with $j\neq 0$, vanish by charge conservation, while the matrix of spin susceptibilities can always be transformed to a diagonal form. In general, there are three spin modes, whose frequencies are found from the equations $1+\delta_jU_{\mathrm{x}}\Pi^0_{ii}=0$, with $\delta_{1}=\delta_2=1$ and $\delta_3=1/2$.
The Green's functions in $\Pi^0_{ij}$ contain the self-energy parts. However, for the special case of
$U_{\mathrm{x}}=$constant, they drop out (see below) and,
after analytic continuation ($iq_0\to \Omega+i\delta$), one obtains
\beq
\label{eq:b}
\Pi^0_{33}(\Omega)=2\nu
\left\langle
 \frac{\Delta_{\vec k_F}^2}{(\Omega+i\delta)^2-\Delta_{\vec k_F}^2}\right\rangle_{\mathrm{FS}},
\eeq
where we have already assumed that SOC is weak in the sense specified above, $\nu$ is the density of states per spin projection,
$\langle \dots \rangle$ denotes
averaging over the Fermi surface (FS),
$\vec k_F=k_F \vec k/k$, and $k_F$ is the Fermi momentum in the absence of SOC. In general,
$\Delta_{\vec k}$ varies from $\Delta_{\min}$ to $\Delta_{\max}$ along the FS. The continuum of inter-subband particle-hole excitations, where $\I\Pi^0_{33}\neq 0$, is confined to the interval $\Delta_{\min}\leq\Omega\leq \Delta_{\max}$. At the boundaries of the continuum, $\R\Pi_{33}^0$ has square root singularities,\cite{SMI} which guarantee a solution of the eigenmode equation $1+U_{\mathrm{x}}\Pi^0_{33}/2 =0$ for $\Omega<\Delta_{\min}$ even at weak coupling. If SOC is isotropic, $\Delta_{\min}=\Delta_{\max}\equiv\Delta$, the two square-root singularities merge into a single pole at $\Omega=\Delta$, the inter-subband continuum shrinks to a single point, and the mode frequency is given by $\Omega_3=\Delta\sqrt{1-u}$, where $u\equiv U_{\mathrm{x}}\nu$.\cite{sm}

Renormalization of the mode frequency by {\em eei} in the SOC case and the lack thereof in the SL case is an important difference, which we discuss now as it will help us to understand the differences in damping later on. This difference occurs because the Greens' functions in the RPA series in Fig.~\ref{fig:1}  include chirality- or spin-dependent shifts in the chemical potential,  which are given by the momentum- and frequency-independent parts of the self-energy.  Each rung of the ladder diagram ($\Pi_{ij}^0$)
contains a difference of the self-energies
\beq
\delta\Sigma_s=\Sigma_s-\Sigma_{-s}
=-U_{\mathrm{x}}\sum_{s'}\int_P \left(B_{s,s'}-B_{-s,s'} \right) g_{s'}(P),\label{se}
\eeq
where $B_{s,s'}$ is the matrix element for the transition $s\to s'$.
Within a given rung, $\delta\Sigma_s$ renormalizes the spin-splitting perturbation, be it SOC or the magnetic field. Since $s=s'$ in the SL case, the self-energy of an electron with given spin is proportional to the number density of electrons with the same spin. Hence, $\delta\Sigma_s$ is proportional to magnetization
[$\delta\Sigma_s=s u\Delta_B/(1-u)$], and each rung of the diagram contains the renormalized Zeeman energy of a quasiparticle,
$\tilde\Delta_B=\Delta_B/(1-u)$. The boundary of the continuum at $q=0$ is shifted from $\Delta_B$ to $\tilde \Delta_B$ (cf. Fig.~\ref{fig:1}, left) as can be seen, {\em e.g.}, from the $11$ component of the rung: $\Pi^0_{11}(\Omega)=-\nu\frac{2\tilde\Delta_B^2}{(\Omega+i\delta)^2-\tilde\Delta_B^2}$. The Larmor theorem is effected via a cancellation between the self-energy and vertex contributions:\cite{ma:1968} when the rung is substituted into the eigenmode equation, the factor of $1-u$ cancels out and the frequency of the mode coincides with bare $\Delta_B$.

The SOC case is different in that chirality, in contrast to spin, is not conserved by {\em eei}, and the sum over $s'$ in Eq.~(\ref{se}) contains both the $s=s'$ and $s'=-s$ terms. For $U_{\mathrm{x}}=\mathrm{const}$, this implies that the self-energy of an electron with given chirality is proportional to the total number density, and thus $\Delta\Sigma_s=0$.\cite{ali_rashba_maslov} The vertex part is, however, non-zero. Therefore,
there is no cancellation between the self-energy and vertex contributions, and the frequencies of the modes are renormalized by {\em eei}. If $U_{\mathrm{x}}\neq \mathrm{const}$, one can show\cite{Raikh,SMII} that $\delta\Sigma_{s}$ does not contain the zeroth angular harmonic
of the interaction (which is why $\delta\Sigma_{s}=0$ for $U_{\mathrm{x}}=\mathrm{const}$), whereas the vertex contribution does. Thus, there is no cancellation between the two contributions in the general case as well.

Damping of collective modes in the region of frequencies and momenta outside the single-particle continuum occurs via
generation of multiple particle-hole pairs, which requires a dynamic interaction, {\em e.g.}, a dynamically screened Coulomb potential. The self-energy of collective modes is depicted diagrammatically in Fig.~\ref{fig:3} {\em a-e}.
The same set of diagrams has been encountered in the analysis of various two-particle correlation functions in the case of an RPA-type interaction.\cite{mirlin,reizer:2000,pepin,avc_dm:2009}
Although the Aslamazov-Larkin (AL) diagrams {\em d} and {\em e} contain two wavy lines, they are of the same order in the bare coupling constant of the theory (the electron charge in our case),  as diagrams {\em a-c}. However,
the contribution of the AL diagrams to damping vanishes within the approximations made in this work.\cite{zak,smIII}

In the case of the SL mode, renormalization of the transverse spin susceptibility ($\chi_{\perp}\sim \chi_{11} + \chi_{22}$) by diagrams {\em a-c} is given by $\delta\chi_{\perp}(Q)=-\mu_B^2\left(\Pi_a+\Pi_b+\Pi_c\right)$, where
\bea
\Pi_a&=&\int_K\left[g_-^2(K)g_+(K+Q)\Sigma_{-}(K) + \left(\Delta_{\vec{k}_F}\rightarrow-\Delta_{\vec{k}_F}\right)\right],\nonumber
\eea
\bea\label{eq:abc-silin}
\Pi_b&=&\int_K\left[g_-^2(K+Q)g_+(K)\Sigma_{-}(K+Q)\right.\nonumber\\
&&\left. +\left(\Delta_{\vec{k}_F}\rightarrow-\Delta_{\vec{k}_F}\right)\right],\nonumber\\
\Pi_c&=&
\int_K \left(\frac{g_-(K+Q)g_+(K)}{iq_0+\Delta_B}\left[\Sigma_+(K)-\Sigma_-(K+Q)\right]\right.\nonumber\\
&&\left. +\left[\Delta_{\vec{k}_F}\rightarrow-\Delta_{\vec{k}_F}\right]\right),
\eea
$\pm$ denote up/down spins,
$Q=(iq_0,0)$, $\Sigma_{\pm}(K)=-\int_P g_{\pm}
(K+P)V_{\text{eff}}(P)$, and $V_{\text{eff}}(P)$ is some dynamic interaction. In the last line of Eq.~(\ref{eq:abc-silin}), we used the identity
\beq g_{\pm}(K+P+Q)g_{\mp}(K+P)=\frac{g_{\pm}(K+P)-g_{\mp}(K+P+Q)}{iq_0\pm \Delta_B}
\label{ident}
\eeq
and integrated over $P$.  Because the denominator in Eq.~(\ref{ident}) does not depend on $P$, this last step produced the same self-energies, $\Sigma_{\pm}$, as in diagrams {\em a} and {\em b}. Adding up the three lines of Eq.~(\ref{eq:abc-silin}), we arrive at
\beq
\label{abc}
\Pi_a+\Pi_b+\Pi_c=\frac{2\Delta_B}{q_0^2+\Delta_B^2}(A_+-A_-),
\eeq
where $A_{\pm} \equiv \int_K g^2_{\pm}(K)\Sigma_{\pm}(K)$. Recalling that $V_{\mathrm{eff}}(P)$ is real on the Matsubara axis and changing the variables as $k_0\rightarrow -k_0$ and $p_0\rightarrow -p_0$, we find that $A_{\pm}=A_{\pm}^*$.  Thus the frequency-independent prefactor in Eq.~(\ref{abc}), $A_+-A_-$,  is real.
Continuing $iq_0$ to the real axis, we see that the imaginary part of $\delta\chi_{\perp}$ comes only from a resonance at the bare Larmor frequency, $\Omega=\Delta_B$, which coincides with the pole of $\chi_{\perp}$ in the RPA approximation. The only effect of the interaction processes represented by diagrams {\em a-c} is thus to renormalize the amplitude of the SL mode without either shifting
its frequency or smearing it.

For the case of SOC, it is convenient to consider renormalization of the out-of-plane spin susceptibility,
$\delta\chi_{33}(Q)=-\mu_B^2
\left(\Pi_a+\Pi_b+\Pi_c\right)$, where now
\begin{widetext}
\bea\label{eq:abc}
\Pi_a&=&\int_K\frac12\left[g_-^2(K)g_+(K+Q)\{\Sigma_{+-}(K)+\Sigma_{-+}(K)\} + g_+^2(K)g_-(K+Q)\{\Sigma_{++}(K)+\Sigma_{--}(K)\}\right],\nonumber\\
\Pi_b&=&\int_K\frac12\left[g_-^2(K+Q)g_+(K)\{\Sigma_{+-}(K+Q)+\Sigma_{-+}(K+Q)\} + g_+^2(K+Q)g_-(K)\{\Sigma_{++}(K+Q)+\Sigma_{--}(K+Q)\}\right],\nonumber\\
\Pi_c&=&-\int_K\int_P\frac12V_{\text{eff}}(P)\left(
 \frac{\mathcal{N}_{-+}\mathcal{M}_{+--}}{iq_0+\Delta_{\vec{k}+\vec{p}}}
+\frac{\mathcal{N}_{+-}\mathcal{M}_{+-+}}{iq_0+\Delta_{\vec{k}+\vec{p}}}
+\frac{\mathcal{N}_{-+}\mathcal{M}_{-++}}{iq_0-\Delta_{\vec{k}+\vec{p}}}
+\frac{\mathcal{N}_{+-}\mathcal{M}_{-+-}}{iq_0-\Delta_{\vec{k}+\vec{p}}}\right),\nonumber\\
\mathcal{M}_{rts}&=&\left[g_r(K+P)-g_t(K+P+Q)\right]
\left[1+s\cos\left(\phi_{\bk}-\phi_{\bk+\bp}\right)\right],\;\mathcal{N}_{rt}=g_r(K)g_t(K+Q).
\eea
\end{widetext}
Here, $r,t,s =\pm $ label the spin-split bands, $\phi_{\vec k}$ depends on the azimuthal angle $\theta_{\vec k}$ of $\vec k$ (and is equal to $\theta_\bk$ for linear Rashba SOC),\cite{smIV} and the partial self-energies are defined as $\Sigma_{rt}(K)=-\int_P g_r(K+P)V_{\text{eff}}(P)\left[1+t\cos(\phi_{\bk}-\phi_{\bk+\bp})\right]$, such  that the total self-energies of the Rashba subbands are  $\Sigma_{+}=\Sigma_{++}+\Sigma_{--}$ and $\Sigma_-=\Sigma_{+-}+\Sigma_{-+}$.
In the last line of Eq.~(\ref{eq:abc}), we only used the identity (\ref{ident}).
In contrast to the SL case, however, the spin-orbit splittings in the denominators of $\Pi_{c}$ depend on
$\vec k+\vec p$, and thus integration over $P$ does not, in general, produce the self-energies. This already tells us that, in general,  the imaginary part of $\delta\chi_{33}$ cannot cancel out between the self-energy and vertex diagrams.

However, there are two realistic approximations, namely, of a long-range interaction ($p\ll k_F$) and of weak SOC ($\Delta_{\bk}\ll E_F$), within which the momentum dependence of $\Delta_{\bk+\bp}$ can be neglected. Assuming that these two conditions are satisfied, the vertex part can again be rewritten in terms of the partial self-energies. Even in this limit, however, there is no complete cancellation between the self-energy and vertex diagrams. Namely, we find that $\Pi_{a}+\Pi_b+\Pi_c$ can be rewritten as $\Pi_{\mathrm{R}} + \Pi_{\text{D}}$, where $\Pi_{\mathrm{R}}=\int_K\frac{\Delta_{\vec k_F}}{q_0^2+\Delta_{\vec k_F}^2}\left\{\Sigma_+(K)g_+^2(K)-\Sigma_{-}(K)g_-^2(K)\right\}$ and
\begin{widetext}
\bea\label{eq:reenter}
\label{sun1}
\Pi_{\text{D}}=-\int_K \frac{\Delta_{\vec k_F}}{q_0^2+\Delta_{\vec k_F}^2}\left\{\left[\Sigma_{--}(K+Q)-\Sigma_{+-}(K)\right]g_-(K)g_+(K+Q) - \left(\Delta_{\vec{k}_F}\rightarrow-\Delta_{\vec{k}_F}\right)\right\}.
\eea
\end{widetext}
The first term, $\Pi_{\mathrm{R}}$, has the same structure as in Eq.~(\ref{abc}); using the same arguments as before, we conclude that $\Pi_{\mathrm{R}}$ does not contribute to damping. In contrast, the second term, $\Pi_{\mathrm{D}}$, does have, in general, an imaginary part at all frequencies, which means damping. To calculate $\Pi_{\mathrm{D}}$ explicitly, one needs to specify the interaction, which we choose to be in the form of a dynamically screened Coulomb potential. Deferring the computational details to Sec.~IV of the Supplementary Material, we quote here only the final result for $\chi_{33}$; near the resonance at $\Omega=\Omega_3$,
\bea\label{eq:final3}
&\chi_{33}^{-1}(\Omega)=(2\nu\mu_B^2)^{-1}A\left[\Omega_3^2-(\Omega+i\Gamma/2)^2\right]\nonumber\\
&A=\left\langle\left(\frac{\Delta_{\vec k_F}\xi_{\bk_F}}{\Delta_{\vec k_F}^2-\Omega_3^2 }\right)^2\right\rangle_{\mathrm{FS}}\left\langle\frac{\Delta^2_{\vec k_F}\xi^2_{\bk_F}}{\Delta_{\vec k_F}^2-\Omega_3^2 }\right\rangle_{\mathrm{FS}}^{-2};&\nonumber\\
&\Gamma= \frac{\omega_{\mathrm{C}}^2}{2E_F}\left(\frac{\Delta_{\bk_F}
}{E_F}\right)^2,
\eea
where $\omega^2_{\mathrm{C}}= r_s^2 E_F^2 \ln r_s^{-1}/12\pi$, $r_s=\sqrt{2}e^2/v_F$ is the coupling constant of the Coulomb interaction, and $\xi^2_{\bk_F}$ is a dimensionless form-factor which depends on the details of SOC; for isotropic SOC, $\xi^2_{\bk_F}=1$.

The damping rate $\Gamma$ has an expected FL form. Notice though that the {\em quasiparticle} damping rate in 2D scales as $\Gamma_{\mathrm{qp}}\propto \Omega^2\ln\Omega$, as opposed to just $\Omega^2$, with a prefactor which does not depend on $r_s$.\cite{narozhny} Being a gauge-invariant quantity, $\Gamma$ contains the {\em differences} of the single-particle self-energies [see Eq.~(\ref{sun1})], while $\Gamma_{\mathrm{qp}}$ is related to the self-energy itself. The infrared singularity in the self-energy, which gives rise to the $\ln\Omega$ factor in $\Gamma_{\mathrm{qp}}$, cancels out in $\Gamma$. As a result, $\Gamma$ is on the order of the {\em transport} decay rate, which is much smaller than $\Gamma_{\mathrm{qp}}$. The FL nature of the result for $\Gamma$ indicates that it would not change substantially if, instead of a FL with Coulomb interaction, we would consider a FL of neutral particles with short-range interaction. The only change would be in $\omega_{\mathrm{C}}^2$ which, for the case of a contact interaction with coupling $U$, should be replaced by $\sim (U\nu E_F)^2$.

Since the frequencies of chiral-spin modes are proportional to the spin-orbit splitting, one might be tempted to conclude that it is better to look for these modes in materials with strong SOC. Our result in Eq.~(\ref{eq:final3}) shows that the advantage of strong SOC has its limits. Indeed, the ratio of the linewidth to the mode frequency, $\gamma\equiv \bar\Gamma/\bar\Delta$, scales as $C\bar\Delta/E_F$, where $\bar\Gamma$ and $\bar\Delta$ are the  appropriate angular averages of $\Gamma$ and $\Delta_{\bk _F}$, correspondingly, and $C$ a dimensionless prefactor.  In a material with sufficiently strong {\em eei}, one should expect that $C\sim 1$. If, in addition,  SOC is also strong ($\bar\Delta\sim E_F$), then $\gamma\sim 1$ and the mode is overdamped. We emphasize that this effect is a unique feature of SOC; in contrast, the SL mode remains undamped (at $q=0$) even if the Zeeman energy becomes comparable to the Fermi energy.

In a particular model of the screened Coulomb potential, the effect of damping appears to be rather weak. For isotropic SOC, Eq.~(\ref{eq:final3}) yields $C=r_s^2\text{ln}r_s^{-1}/12\pi$. Using parameters for an InGaAs/InAlAs quantum well, we then find $\gamma\approx 2\times10^{-3}\bar\Delta /E_F$ for an electron number density of $1.6\times10^{12}$ cm$^{-2}$. On the other hand, chiral-spin waves are also damped by disorder via the Dyakonov-Perel' mechanism.\cite{shekhter} For a mobility of $2\times10^{5}$ cm$^2$/V$\cdot$ s, damping due to disorder  is stronger than that by {\em eei} by a factor of ten.  One should not forget, however, that Eq.~(\ref{eq:final3}) is valid only for $r_s\ll 1$ and the actual numbers may differ from quoted above as $r_s$ increases. Although damping from disorder
appears to be the dominant effect in solid-state systems, damping due to
interaction should be dominant in (fermionic) cold-atom systems with synthetic SOC,\cite{ca1,ca2} which have virtually no disorder. In this case, the interaction is short-ranged but, as we have already mentioned,
this should only affect the prefactor in Eq.~(\ref{eq:final3}).

In conclusion, we showed that {\em eei}  in the presence of SOC not only gives rise to a new type of collective modes but also leads to their damping. This damping occurs even at $q=0$ and its rate scales as the square of the spin-orbit splitting. This effect occurs because neither of the three components of magnetization is a good quantum number in the presence of SOC.

We would like to thank M. Imran and V. Zyuzin for useful discussions. SM is a Dirac Post-Doctoral Fellow at the National High Magnetic Field Laboratory, which is supported by the National Science Foundation via Cooperative agreement No. DMR-1157490, the State of Florida, and the U.S. Department of Energy. DLM acknowledges support from the National Science Foundation via grant NSF DMR-1308972.

\newpage
\begin{widetext}

\section*{Supplementary Material}

\section{Threshold singularities in Equation (4) of the main text (MT) for
generic spin-orbit coupling (SOC)}

For a generic (but still weak compared to the Fermi energy SOC) the spin-orbit band splitting, $\Delta_{\bk k_F}$, is anisotropic. Suppose that $\Delta_{\bk k_F}$
varies from $\Delta_{\text{min}}$ to $\Delta_{\text{max}}$ along the Fermi surface (FS). Quite generally,
the angular dependence of $\Delta_{\bk k_F}$ near the extremal points is $\Delta_{\text{min}}\approx \Delta_{-}+\beta^2_{+}\theta^2$ and $\Delta_{\text{max}}\approx \Delta_{+}-\beta^2_{-}\theta^2$. Near each of these extremal points, the diverging part of the angular integral in Eq.~(4) of MT can be written as
\bea
\left\langle\frac{\Delta_{\bk_F}^2}{\Omega^2-\Delta_{\bk_F}^2}\right\rangle_{\mathrm{FS}} &\approx &
\left\{
\begin{array}{cll}
\frac 12\int\frac{d\theta}{2\pi}~
\frac{\Delta_+}{(\Omega-\Delta_++\beta_+^2\theta^2)}\propto\frac{1}{\sqrt{\Omega-\Delta_+}},~~~\text{near maximum},\nonumber\\
\frac12\int\frac{d\theta}{2\pi}~
\frac{\Delta_+}{(\Omega-\Delta_--\beta_-^2\theta^2)}\propto-\frac{1}{\sqrt{\Delta_--\Omega}},~~~\text{near minimum}.
\end{array}
\right.
 \label{eq:app1}
\eea
Note that the angular average is imaginary when $\Delta_-<\Omega<\Delta_+$, which correspond to the continuum of inter-subband particle-hole excitations. The singularity in the real part of the susceptibility at the lower end of the continuum ensures that the collective modes exist even for infinitesimally weak electron-electron interaction ({\em eei}). For an isotropic SOC, {\em e.g.}, either for Rashba or Dresselhaus SOC, $\Delta_+=\Delta_-\equiv \Delta$ and
the two square-root branch cuts merge into a simple pole $1/(\Omega-\Delta)$.

\section{Static part of the self-energy for momentum-dependent interaction}
\label{sec:Sigma}
In the MT, we argued that the difference of the self-energies of chiral subbands does not contain the zeroth angular harmonic of the interaction potential. In this section, we prove this statement. The self-energy of a subband with chirality $s=\pm 1$ due to {\em eei} via a static but momentum-dependent potential, $U_{|\bq|}$, is equal to
\beq\label{eq:self energy} \Sigma_s(K)=-\frac12\int_{K'} \left
\{g_+(K')+g_-(K')+s\left[g_+(K')-g_-(K')\right]\cos\phi_{\bk\bk'}\right\}U_{|\bk-\bkp|}.
\eeq
For the case of linear Rashba SOC, $\phi_{\bk \bk'}$ is the angle between $\vec{k}$ and $\vec{k'}$ ($\equiv \theta_{\bk\bk'}$); for a general case, $\phi_{\bk \bk'}$ is some function of $\theta_{\bk\bk'}$
such that $\int\cos\phi_{\bk\bk'} d\theta_{\bk\bk'} =0$. We first consider the case of Rashba SOC in some detail and then point out the qualitative features that remain the same for arbitrary SOC.
In addition, we also assume that Rashba SOC is weak ($\lambda\ll v_F$), in which case the spin-orbit splitting can be approximated by its value projected on the FS in the absence
of SOC: $\Delta\approx 2\lambda k_F$. The difference of the self-energies of the chiral subbands, $\delta\Sigma_s\equiv \Sigma_s-\Sigma_{-s}$, is now given by
\bea\label{eq:delta sigma}
\delta\Sigma_s(K)&=&-s \int_{\vec{k}'} \left[n_F\left(\varepsilon_{\bk'}+\frac{\Delta}{2}\right)-n_F\left(\varepsilon_{\bk'}-\frac{\Delta}{2}\right) \right]\cos\theta_{\bk\bk'}U_{|\bk-\bk'|},
\eea
where $n_F(\epsilon)$ is the Fermi function.
Setting $T=0$ and expanding the Fermi functions in $\Delta$, we find
\beq
\delta\Sigma_s(K)=s\nu\Delta\int\frac{d\theta_{\bk\bk'}}{2\pi} \cos\theta_{\bk\bk'}U_{|\bk-\bk'_F|},
\eeq
where $\bk'_F=k_F \bk'/\bk$. To find the frequency of the collective mode, one needs to project $\bk$ on the FS upon which  $\delta\Sigma_s$ becomes proportional to the first harmonic
of the interaction potential.

A similar argument can be made for an arbitrary (but still weak) SOC. The Green's function in Eq.~(2) of MT can be separated into symmetric (S) and asymmetric (A) parts: $G=G^{\mathrm{S}}+G^{\mathrm{A}}$. Any crystalline plane has at least a $C_2$ symmetry upon which $\bk\to -\bk$. This  guarantees that $\Delta_{-\bk}=\Delta_{\bk}$ and thus the subband Green's functions,
$g_{\pm}$, are even on $\bk\to -\bk$. Then $G^{\mathrm{S}}$ is even while $G^{\mathrm{A}}$ is odd on this operation. Consider the self-energy in the spin basis (a $2\times 2$ matrix)
\beq
\Sigma(K)=-\int_{K'} U_{|\bk-\bk'|} G(K'),
\eeq
which can be also decomposed into symmetric and asymmetric parts: $\Sigma=\Sigma^{\mathrm{S}}+\Sigma^{\mathrm{A}}$.
The zeroth angular harmonic of the interaction potential, $U^{\{0\}}$,  survives only in the part of the integral associated with $G^{\mathrm{S}}$ and enters the diagonal elements of $\Sigma$ as $-U^{\{0\}}n$, where $n$ is the total number density. When $\Sigma$ is transformed to the (diagonal) chiral basis, the  $-U^{\{0\}}n$ terms remain on the diagonal and cancel out in the difference $\Sigma_{+}-\Sigma_{-}$.

\section{Aslamazov-Larkin (AL) diagrams}
In the presence of a dynamic and long range interaction, the AL diagrams are of the same order as the self energy and vertex diagrams (See Refs. 29-32 of MT). In the absence of  SOC and for a spin-invariant interaction, however, AL diagrams for the spin susceptibility vanish for a trivial reason: Tr[$\sigma_i {G} {G}{G}$]$=0$, ($i=1,2,3$) because ${G} \propto\sigma_0$. In the presence of the magnetic field (along the $3$-axis),  ${G}= a\sigma_0 + b\sigma_3$. The relevant quantity in this case in the transverse spin susceptibility, which contains the trace Tr[$\sigma_i {G}{G}{G}$] with $i=1,2$. Again, this trace is equal to zero. The vanishing of the AL diagrams
in both these cases is a consequence of conservation of either total spin (in the former case) or its component along the magnetic field (in the latter case). In the presence of SOC, this reasoning does not apply because ${G}$ now has off-diagonal elements.  However, some general statements about the AL diagrams can still be made for $q=0$ case.

\begin{figure}[htp]
\includegraphics[width=0.7\columnwidth]{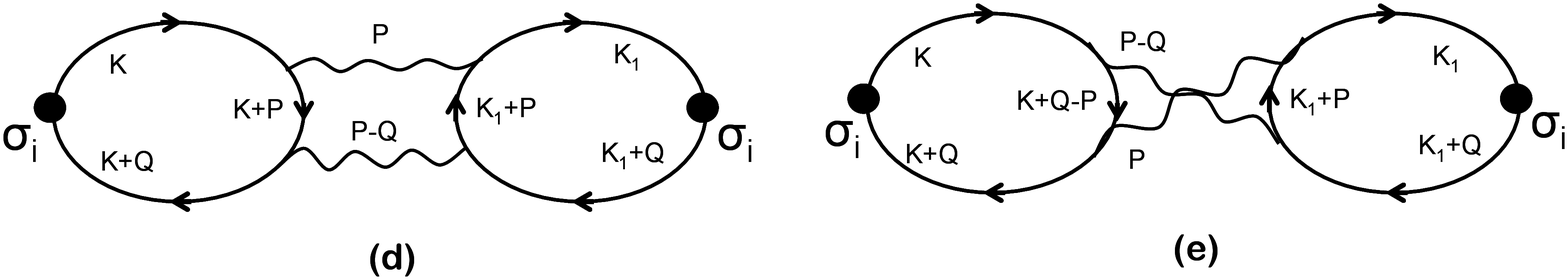}
\caption{
\label{fig:AL} Detailed AL diagrams from Fig. 2 of MT}
\end{figure}

For convenience, we present the AL diagrams from Fig. 2, {\em d} and {\em e} of MT here in Fig.~\ref{fig:AL}. It suffices to consider only the diagonal components of the spin susceptibility,
$\chi_{ii}$, with $i=1\dots 3$, in which case the diagrams read
\bea\label{eQ:AL1} \Pi^d_{ii}&=&\int_{P}\int_K \text{Tr}\left[ \sigma_i{G}(K){G}(K+P){G}(K+Q)\right]  V_PV_{P-Q}\int_{K_1}\text{Tr}\left[\sigma_i{G}(K_1+Q){G}(K_1+P){G}(K_1)\right]  , \nonumber\\
\Pi_{ii}^e&=&\int_{P}\int_K \text{Tr}\left[\sigma_i {G}(K+Q){G}(K+Q-P){G}(K) \right]  V_PV_{P-Q}\int_{K_1}\text{Tr}\left[\sigma_i{G}(K_1+Q){G}(K_1+P){G}(K_1)\right].  \nonumber\\
\eea
Here, $Q$ has only the temporal component: $Q=(iq_0,\vec 0)$. We can define
\bea\label{eq:4func}
h_i(P,Q)\equiv\int_K\text{Tr}\left[\sigma_i{G}_{K+Q}{G}_{K+P}{G}_K\right],\nonumber\\
\tilde{h}_i(P,Q)\equiv\int_K\text{Tr}\left[\sigma_i{G}_{K}{G}_{K+P}{G}_{K+Q}\right],\nonumber\\
f_i(P,Q)\equiv\int_K\text{Tr}\left[\sigma_i{G}_{K+Q}{G}_{K+Q-P}{G}_K\right].
\eea
Using the matrix structure of ${G}(K)$, we observe that\beq\label{eq:AL2}h_{1,2}(P,Q)=+\tilde{h}_{1,2}(P,Q);~ h_3(P,Q)=-\tilde{h}_3(P,Q)\; \text{and}\;\;f_i(P,Q)=-\tilde h_i^*(P,Q)~~\text{for}~~ i=1,2,3.
\eeq
With the help of these properties, the AL diagrams can now be written as
\bea
\Pi_{ii}^d&=&\int_{P} h_i(P,Q)\tilde{h}_i(P,Q)  V_PV_{P-Q}= \int_{P} h_i^2(P,Q)  V_PV_{P-Q}~\text{for}~i=1,2~\text{and}~\nonumber\\
&=&-\int_{P} h_i^2(P,Q)  V_PV_{P-Q}~\text{for}~i=3,\label{eQ:AL5}\\
\Pi_{ii}^e
&=&\int_{P} f(P,Q)\tilde{h}_i(P,Q)  V_PV_{P-Q}=-\int_{P}|h_i(P,Q)|^2 V_PV_{Q-P},
\eea
and thus their sum is reduced to
\bea\label{eQ:AL6}
\Pi^d_{ii}+\Pi^e_{ii}
&=&\int_{P} \left(h_i^2(P,Q) - |h_i(P,Q)|^2\right) V_PV_{P-Q}, ~~~\text{for}~~i=1,2~~\text{and}\nonumber\\
&=&-\int_{P}
\left(h_i^2(P,Q) + |h_i(P,Q)|^2\right) V_PV_{P-Q}, ~~~\text{for}~~i=3.
\eea

We now study two particular cases--that of Rashba SOC and of combined Rashba-Dresselhaus SOC--
and show that the AL diagrams do not contribute to damping in either of these two cases.

\emph{Rashba SOC}: For $i=3$,
\beq
\label{h3}
h_3(P,Q)=-\frac i4\sum_{a,b,c=\pm} b(a-c)\int_K \sin(\theta_{\vec k}-\theta_{\vec k+\vec p})g_a(K+Q)g_b(K+P)g_c(K).
\eeq
Since $Q$ has no spatial component, we can choose the direction of $\bp$ as a reference. On reflecting vector $\bk$ about this direction, both $\theta_\bk$ and $\theta_{\bk+\bp}$ change signs and so does the factor of $\sin(\dots)$. On the other hand,  since the energy spectrum is isotropic, each of the three subband Green's function in the formula above is a function of the magnitude of the corresponding momentum. It is then easy to see that all the three Green's function are even on reflection $k_1\to k_1$, $k_2\to -k_2$. Therefore, the integral over $\theta_{\bk}$ vanishes, and the AL diagrams give no  contribution to
$\chi_{33}$.

The same reasoning applies for $i=1$. In this case,
\beq
\label{h1}
h_1(P,Q)=\frac14\sum_{a,b,c=\pm}\int_K\left[(a+c)\sin\theta_{\bk}+abc\sin(2\theta_{\bk}+\theta_{\bk+\bp})
+b\sin\theta_{\bk+\bp}\right]g_a(K+Q)g_b(K+P)g_c(K)
\eeq
contains again an angular factor, which is odd on reflection, and a combination of the Green's functions, which is even on reflection. Therefore,  $h_1(P,Q)=0$ and there is no contribution to $\chi_{11}$ from the AL diagrams either.

The same argument cannot be used for $i=2$, because the angular factor in
\beq
\label{h2}
h_2(P,Q)=-\frac14\sum_{a,b,c=\pm}\int_K\left[(a+c)\cos\theta_{\bk}+abc\cos\left(2\theta_{\bk}+\theta_{\bk+\bp}\right)
+b\cos\theta_{\bk+\bp})\right]g_a(K+Q)g_b(K+P)g_c(K).
\eeq
is even on reflection. Nevertheless,  in-plane rotational symmetry of the Rashba Hamiltonian ensures that $\chi_{11}=\chi_{22}$. Therefore, one does not need to consider $\chi_{22}$ separately: if there is no damping of resonance associated with $\chi_{11}$, the same is true for $\chi_{22}$.

\emph{Combined Rashba and Dresselhaus SOC}: In the presence of both Rashba and Dresselhaus SOC, the only change in Eqs.~((\ref{h3}-\ref{h2})) is that $\theta_{\bk}$ needs to replaced by $\phi_{\bk}$, as defined after Eq.~(\ref{eq:self energy}). It is convenient to choose the Cartesian system rotated by $\pi/4$ compared to the conventional one. In such a system, the combined Rashba plus Dresselhaus Hamiltonian is described by in Eq.~(1) of MT with $\lambda\vec f (k)=\left([\beta+\alpha]k_2,[\beta-\alpha]k_1,0\right)$ and
\bea\label{eq:app2}
\cos\phi_{\bk}&=&-\frac{(\alpha-\beta) \cos\theta_\bk}{\Lambda_\bk},\nonumber\\
\sin\phi_\bk&=&-\frac{(\beta+\alpha)\sin\theta_\bk}{\Lambda_\bk},\eea
where $\Lambda_\bk=\sqrt{\alpha^2+\beta^2-2\alpha\beta\cos2\theta_\bk}$. Note that $\phi_{\bk}=\tan^{-1}\left(\frac{\alpha+\beta}{\alpha-\beta}\tan\theta_{\vec k}\right)$.
The spin-orbit splitting $\Delta_{\bk_F}=2k_F\Lambda_{\bk}$ is even on reflection
about the $k_1$-axis.  As discussed in MT one can approximate $\Delta_{\vec k + \vec p}$ by $\Delta_{\vec{k}_F}$  for a long-range interaction; therefore, the products of the Green's functions are still even on reflection.  Because of the  relation between $\phi_{\vec k}$ and $\theta_{\vec k}$, the same symmetries  that led to  the  vanishing of $h_{1}$ and $h_{3}$  in the case with only Rashba SOC  ensure that $h_{1}$ and $h_{3}$ vanish is  the case when both Rashba and Dresselhaus SOC are present  as well.

\section{Derivation of Equation (11) of the Main Text \label{suppl:damping}}
In this section, we present the evaluation of $\Pi_{\text{D}}$ from Eq.~(10) of the MT. For convenience, we present this equation below in an expanded form:
\bea
\label{sun1}
\Pi_{\text{D}}=-\int_K \frac{\Delta_{\vec k_F}}{q_0^2+\Delta_{\vec k_F}^2}\left\{\left[\Sigma_{--}(K+Q)-\Sigma_{+-}(K)\right]g_-(K)g_+(K+Q) - \left[\Sigma_{+-}(K+Q)-\Sigma_{--}(K)\right]g_+(K)g_-(K+Q)\right\},\nn\\
\eea
where
$\Sigma_{ab}(K)=-\int_P g_a(K+P)V_{\text{eff}}(P)\left[1+b\cos(\phi_{\bk}-\phi_{\bk+\bp})\right]$ and the angle $\phi_{\bk}$ is defined after Eq.~(\ref{eq:self energy}). Since $\Pi_{\text{D}}$ is already proportional to $\Delta_{\vec k_F}$, we can neglect the effect of SOC on the interaction potential and  take it to be a dynamically screened Coulomb potential in 2D:
\beq
\label{eq:scrC}
V_{\mathrm{eff}}(P)=\frac{2\pi e^2}{p+\kappa\left(1-\frac{|p_0|}{\sqrt{p_0^2+v_F^2p^2}}\right)},
\eeq
where $\kappa= \sqrt{2}r_s k_F$ is the inverse Thomas-Fermi screening length and $r_s =\sqrt{2}e^2/v_F$. The momentum transfer $\vec p$ is decomposed into two components: along the normal to the FS at point $\vec k$ ($p_{||}$) and along the tangent ($p_\perp$). We assume and then verify that the quasiparticle interaction is determined by processes with typical $p_{||}$ on the order of $p_0/v_F$, whereas typical $p_\perp$ are on the order of $\kappa$. In turn, typical $p_0$ are on the order of the external frequency which, in our case, is fixed by spin-orbit splitting, $\bar\Delta$, averaged over the FS. We choose to work in a realistic regime of $\bar\Delta\ll v_F\kappa\ll E_F$. [Although the form of the potential in Eq.~(\ref{eq:scrC}) is, strictly speaking,  valid for $r_s\ll 1$ which implies that $\kappa\ll k_F$, these strong inequalities are never satisfied in real materials. Therefore, the interval of energies in between $v_F\kappa$ and $E_F$ is never wide enough to consider a possibility of $\bar\Delta$ being within this interval.] Since \beq
|p_{||}|\sim |p_0|/v_F\ll |p_\perp|\label{ineq}
\eeq in this regime, we can simplify the Landau-damping term in Eq.~(\ref{eq:scrC}) to $|p_0|/v_F|p_\perp|$ and also expand the denominator of this equation in this parameter. Subtracting off the static screened potential which gives no contribution to damping, we obtain for the dynamical ($p_0$-dependent) part of the interaction
\beq\label{eq:dyn_2}
U^{\text{dyn}}_{\text{eff}}(P)=\left(\frac{2\pi e^2}{|p_\perp|+\kappa}\right)^2
\frac{\nu}{v_F}\frac{|p_0|}{|p_\perp|}.\eeq
Thanks to the inequality (\ref{ineq}), the integrals over the fermionic ($K$)
and bosonic ($P$) momenta become separable. Let us pick one of the terms
in Eq.~(\ref{sun1}): \bea\label{eq:step1}
\int_K \Sigma_{--}(K+Q)g_-(K)g_+(K+Q)=-\int_K \int_P U^{\text{dyn}}_{\text{eff}}g_-(K+Q+P)g_-(K)g_+(K+Q)\left[1-\cos\left(\phi_{\vec k}-\phi_{\vec k
+\vec p}\right)\right].
\eea
Since it is expected that typical $|p_\perp|\sim\kappa\ll k_F$ while $\vec k$ is expected to be near the FS, the angular factor can be expanded as $1-\cos(\phi_{\vec k}-\phi_{\vec k +\vec p}) \approx\xi^2
_{\bk}\frac{p_\perp^2}{2k_F^2}$, where $\xi_{\bk}$ is a form-factor that depends on the details of the SOC.

For example, in the case of linear Rashba SOC, $\phi_{\vec k}$ coincides with the angle $\theta_{\bk}$ that $\vec k$ makes with, e.g., the $1$-axis. Also, due to rotational symmetry, the result of the integral over $P$ does not depend on the direction of $\bk$ and thus we can choose $\theta_{\bk}=0$ in $1-\cos(\theta_{\vec k}-\theta_{\vec k +\vec p})$. An expansion of $1-\cos(\theta_{\vec k +\vec p})$ then yields $\frac{p_\perp^2}{2k_F^2}$, which means that $\xi_{\bk}=1$ for a system with linear Rashba SOC.
Notice that the condition $|p_{||}|\ll |p_{\perp}|$ implies that $\bp$ is almost perpendicular to $\bk$, i.e., that $|\theta_{\bp}-\theta_{\bk}|\approx\pi/2$.

In the presence of both Rashba and Dresselhaus SOC, $\phi_{\bk}$ is given by Eq.~(\ref{eq:app2}). Expanding in  $p$ and using that $|\theta_\bp-\theta_\bk|\approx\pi/2$, we get
\bea\label{eq:app3}
1-\cos(\phi_{\vec k}-\phi_{\vec k +\vec p})&\approx& \frac{p_\perp^2}{2k_F^2}\left[1
+\left(\frac{\gamma}{1-\gamma\cos2\theta_
\bk}\right)^2\right], ~~\text{where}~~\gamma=\frac{2\alpha\beta}{\alpha^2+\beta^2}<1
\eea
Thus $\xi^2_\bk=1+\left(\frac{\gamma}{1-\gamma\cos2\theta_\bk}\right)^2$ for this system. We avoid the special case of  $|\alpha|=|\beta|$, when $\gamma=1$ and $\xi_{\bk}$ has a pole: it can be shown that  collective modes are completely covered by the continuum in this case.\cite{imran}

Next, we write $g_-(K)g_+(K+Q)$ as $[g_-(K)-g_+(K+Q)]/(iq_0-\Delta_{\vec k})$ and integrate the products of two Green's functions over $\e_{\vec k}$, $k_0$ and $p_{||}$; for one of the products, we obtain
\bea\label{eq:temp step}
&&\int\frac{dp_{\parallel}}{2\pi}\int\frac{dk_0}{2\pi}\int d\varepsilon_{\vec k} g_a(K+P)g_b(K)=\int\frac{dp_{\parallel}}{2\pi} \frac{ip_0}{ip_0-v_Fp_{\parallel} - \frac{a-b}{2}\Delta_{\vec k}}=\frac{|p_0|}{2v_F}.
\eea
The rest of the integrals are evaluated in the similar manner. Collecting all terms, we arrive at
\bea\label{eq:leftover}
\Pi_{\mathrm{D}}=
-\left\langle\left(\frac{\Delta_{\vec k_F}\xi_{\bk_F}}{q_0^2+\Delta^2_{\vec k_F}}\right)^2\right\rangle_{\vec k}\frac{2\nu}{v_F k_F^2}\int_0^{k_F}\frac{dp_{\perp}}{2\pi}{p^2_{\perp}}\int_0^\infty \frac{dp_0}{2\pi}
U_{\text{eff}}^{\text{dyn}}(P)\left\{|q_0+p_0|+|q_0-p_0|-2|p_0|\right\}.\eea
The factor in $\{\dots \}$ constraints the range of integration over  $p_0$ to $(0,q_0)$. This leads to
\bea\label{eq:formal}
\Pi_D&=&-\nu\left\langle\left(\frac{\Delta_{\vec k}\xi_{\vec k_F}
}{q_0^2+\Delta_{\vec k}^2}\right)^2\right\rangle_{\vec k}\frac{ r_s^2}{2\pi E_F}\int_0^{q_0} dp_0(q_0-p_0)p_0\int_{0}^{k_F} d p_\perp \frac{p_\perp}{(p_\perp+\kappa)^2}
\eea
To logarithmic accuracy, the integral over $p_\perp$ gives $\ln k_F/\kappa\approx -\ln r_s$ while the integral over $p_0$ gives $q_0^3/6$. Note that typical $p_0\sim q_0$ while typical $p_{\perp}\sim\kappa$ (in the logarithmic sense), and thus our initial assumption is verified.

After analytic continuation, we get \bea\label{eq:final result22}
\Pi_D=-i\nu\left\langle\left(\frac{\Delta_{\vec k_F}\xi_{k_F}}{\Omega^2-\Delta^2_{\vec k_F}}\right)^2\right\rangle_{{\vec k}_F}\omega_C^2\left(\frac{\Omega}{E_F}\right)^3,
\eea
where $\omega^2_C=r_s^2 E_F^2 \ln r_s^{-1}/12\pi$ and $r_s=\sqrt{2}e^2/v_F$.  This is the correction to $\Pi^0_{33}$ that is responsible for damping. Recalling that the physical susceptibility is given by $\chi_{33}=-\mu_B^2\Pi_{33}/(1+\frac{U_{\text{x}}}{2}\Pi_{33})$ $\Pi_{33}$ differs from $\Pi_{33}^0$ due to the damping correction. To account for this correction, we may write
\beq\label{eq:w}
\Pi_{33}=2\nu\left\langle\frac{\Delta_{\vec k_F}^2\xi^2_{\vec k_F}}{\Delta_{\vec k_F}^2-\left(\Omega+i\Gamma/2\right)^2}\right\rangle_{\text{FS}},
\eeq
where $\Gamma = \frac{\omega_C^2}{2E_F}\left(\frac{\Delta_{\vec k_F}}{E_F}\right)^2$. Using the fact that $\Omega_3$ is the pole in $\chi_{33}$ when $\Gamma=0$, we can eliminate the $U_{\text{x}}$ dependence. Writing $\Delta_{\vec k_F}^2-\left(\Omega+i\Gamma/2\right)^2$ as $\Delta_{\vec k_F}^2-\Omega_3^2+\Omega_3^2-\left(\Omega+i\Gamma/2\right)^2$ and expanding in $\Omega_3^2-\left(\Omega+i\Gamma/2\right)^2$ near the resonance, we obtain Eq.~(11) of MT.
\end{widetext}


\begin{thebibliography}{5}
\bibitem{footnote_eei} While we will be using the term ``electron-electron interaction" throughout the paper, it is to be understood that our results either do not depend on the particular form of the interaction or can be extended to other forms, e.g., to a short-range interaction between neutral atoms, with minimal effort.
\end{thebibliography}
\end{document}